\documentclass[a4paper,fleqn,usenatbib]{mnras}

\usepackage{newtxtext,newtxmath}

\usepackage[T1]{fontenc}
\usepackage{ae,aecompl}

\usepackage{graphicx}   
\usepackage{amsmath}    
\usepackage{amssymb}    
\usepackage{longtable}
\usepackage{multirow}
\usepackage{lipsum}
\usepackage{rotating}
\usepackage{bm}
\usepackage{color}
\usepackage[T1]{fontenc}
\usepackage{ae,aecompl}
\usepackage{makecell}

\usepackage[labelformat=parens,labelsep=quad,skip=3pt]{caption}

\title[]{Probing cosmic acceleration by strong gravitational lensing systems}

\author[Z. L. Tu, J. Hu and F. Y. Wang]{
    Z. L. Tu,$^{1}$
    J. Hu,$^{1}$
    F. Y. Wang$^{1,2}$\thanks{E-mail: fayinwang@nju.edu.cn (NJU)}
    \\
    $^{1}$School of Astronomy and Space Science, Nanjing University, Nanjing 210093, China\\
    $^{2}$Key Laboratory of Modern Astronomy and Astrophysics (Nanjing University), Ministry of Education, Nanjing 210093, China
}

\date{Accepted XXX. Received YYY; in original form ZZZ}

\pubyear{2018}

\begin{document}
    \label{firstpage}
    \pagerange{\pageref{firstpage}--\pageref{lastpage}}
    \maketitle

\begin{abstract}
        Recently, some divergent conclusions about cosmic acceleration were
        obtained using type Ia supernovae (SNe Ia), with opposite
        assumptions on the intrinsic luminosity evolution. In this paper, we
        use strong gravitational lensing systems to probe the cosmic
        acceleration. Since the theory of strong gravitational lensing is
        established certainly, and the Einstein radius is determined by
        stable cosmic geometry. We study two cosmological models, $\Lambda$CDM and power-law models,
        through 152 strong gravitational lensing systems, incorporating with
        30 Hubble parameters $H(z)$ and \textbf{11} baryon acoustic oscillation (BAO)
        measurements. Bayesian evidence are introduced to make
        a one-on-one comparison between cosmological models.
        Basing on Bayes factors $\ln B$ of flat $\Lambda$CDM versus power-law and $R_{h}=ct$ models
         are $\ln B>5$,
        we find that the flat $\Lambda$CDM is strongly supported by the combination of the datasets. Namely, an accelerating cosmology with non power-law expansion is preferred by our numeration.
\end{abstract}

\begin{keywords}
    cosmology: theory - gravitational lensing: strong - dark energy
\end{keywords}

\section{Introduction}
As a standard cosmological model, the $\Lambda$CDM is widely
    accepted basing on some remarkable observations, including type Ia
    supernovae (SNe Ia) \citep{1998Riess, 1999Perlmutter}, cosmic
    microwave background (CMB)
    \citep{2011Bennett,2013Bennett,2014Planck,2016Planck}, baryon
    acoustic oscillations (BAOs) \citep{2005Eisenstein} and
    gamma-ray bursts \citep{2015Wang}. But many imperfections of
    $\Lambda$CDM are also needed to be faced, including the fine tuning
    problem and the cosmic coincidence problem \citep{1989Weinberg,
        1999Zlatev}. Additionally, the capability of using SNe Ia to prove
    cosmic acceleration is under doubt recently \citep{2016Nielsen,
        2016Shariff}.

    In a bunch of cosmological models, power-law cosmology was proposed
    with an assumption that the scale factor varies as $a(t)\propto
    t^{n}$ \citep{1997Dolgov, 2014Dolgov}. Many works have devoted into
    the study of power-law cosmology. \citet{2007Melia} and
    \citet{2012Melia} made a special situation of this model, where a
    non-accelerated cosmology was considered as $a(t)\propto t$. Namely,
    the cosmic horizon $R_{ h}$ equals to the light-travel time
    distance. By using different kinds of observations, this
    non-accelerated power-law cosmology was found to be preferred
    \citep{2013Melia,2015MeliaA,2015Melia, 2017Tutusaus}. However this
    model was also perceived to be against to SNe Ia and BAO data
    \citep{2012Bilicki,2014Dolgov,2015Shafer,2017Haridasu}. Progress has
    been made in probing power-law cosmology
    \citep{2005Sethi,2008Dev,2008ZhuA,2014Dolgov,2014Yu}. Nevertheless,
    many researches also questioned the possibility of replacing
    $\Lambda$CDM by power-law cosmology
    \citep{2012Kumar,2015Cardenas,2015Rani,2015Yuan,2016Tutu,2017Haridasu,2017Lin}.

    Treating SNe Ia as standard candles is broadly used in probing
    cosmology, and this method is required to be deliberately considered
    according to some literature recently. \citet{2017Tutusaus} found
    that SNe Ia are consistent with  a non-accelerated universe with
    assumption that supernovae intrinsic luminosity depends on the
    redshift. Considering the independence of supernovae intrinsic
    luminosity on redshift, \cite{2017Lin} got a contradictory
    conclusion with \citet{2017Tutusaus}. \citet{2016Nielsen} presented
    an improved maximum-likelihood procedure to reanalyze the SNe Ia
    dataset with improvement of precision and scale. They found the
    marginal evidence of an accelerated expansion cosmology which was
    widely accepted before. A new method called BAyesian HierArchical
    Modeling for the Analysis of Supernova cosmology (BAHAMAS) was
    introduced by \citet{2016Shariff}, and they found a sharp dropping
    of the color correction parameter value $\beta$ with redshift. Color
    correction parameter is roughly constrained over all redshift
    through empirical period-luminosity relation. However, some voices
    \citep{2016Rubin,2016Ringermacher} have grown to demonstrate an
    accelerated cosmology and pointed some inappropriatenesses in
    \citet{2016Nielsen}. While using SNe Ia, a set of complex parameters
    are involved to be constrained by empirical period-luminosity
    relation of Cepheid variables \citep{1993Phillips}. Several
    uncertainties of this relation are also needed to be prudently
    studied, e.g., the dependence of absolute B band magnitude $M_{
        B}$ on redshift \citep{2005Holz}, the effects of systematic errors
    \citep{2010Freedman,2012Ruiz}, and poor uniformity of SNe Ia in
    various galaxy environment \citep{2010Gilfanov}. When probing cosmic
    acceleration, all the considerations above trigger us to use other
    observations. We choose strong gravitational lensing systems which
    are purely geometrically effected and endorsed by certain theory.

\citet{1979Walsh} discovered the first strong gravitational lensing
    Q0957+561. Decades years later, strong gravitational lensing systems
    are widely used in cosmology
    \citep{2008Zhu,2011Wang,2012Cao,2014Wei,2016Liao,2018Leaf,2018aYu,2018WangY}. Basing
    on the Einstein radius formulation, the Hubble constant is
    eliminated through distance ratio. This may avoid some undiscovered
    relationship between Hubble constant and other observed values,
    e.g., the absolute magnitude $M_{ B}$ of SNe Ia. Considering that
    the strong gravitational lensing systems may show of lacking
    capability to constrain $\Omega_{ m }$ for $\Lambda$CDM
    \citep{2010Biesiada}, other kinds of data are combined to constrain
    parameters precisely in this paper.

    In our work, we focus on using strong gravitational lensing systems,
    Hubble parameters $H(z)$ and BAOs to compare cosmological
    models through Markov chain Monte Carlo (MCMC) method. We also pick a
    favorable cosmological model quantitatively through Bayesian evidence.
    \newline

    This paper is organized as follows. Two cosmological models will be
    briefly introduced in section \ref{sec:Model}. All kinds of data and
    methodology are given in section \ref{sec:samples}. The results of
    MCMC simulation are presented in section \ref{sec:result}.
    Conclusions are given in section
    \ref{sec:discuss}.

    \section{Cosmological Models}\label{sec:Model}
    In this section, we briefly present two cosmological models in our analysis, including
    $\Lambda$CDM and power-law model.

\subsection{$\Lambda$CDM model}
    With an assumption of isotropy and homogeneity of the universe, the Hubble parameter can be derived as
    \begin{equation}
    \label{equ:Hza}
    \left(\frac{H}{H_{0}}\right)^{2}=\Omega_{r}a^{-4}+\Omega_{m}a^{-3}+\Omega_{k}a^{-2}+\Omega_{\Lambda},
    \end{equation}
    where $\Omega_{r}$, $\Omega_{\Lambda}$, $\Omega_{m}$ and $\Omega_{k}$ represent radiation, dark energy, matter and curvature of the universe, respectively. $\Omega_{r}$ is fixed as $\Omega_{r}=0$ in our analysis, basing on the radiation of universe is observed to be negligible in present day. $\Omega_{m}+\Omega_{k}+\Omega_{\Lambda}=1 $ represent the total density which is dimensionless.
    Using $a=(1+z)^{-1}$, eq.(\ref{equ:Hza}) can be derived into
    \begin{equation} \label{equ:HOM}
    \begin{aligned}
    H(z)=H_{ 0 }[\Omega _{ m }(1+z)^{ 3 }+\Omega _{ k }(1+z)^{ 2 }+\Omega _{ \Lambda  }]^{ 1/2 }.
    \end{aligned}
    \end{equation}
    We consider a flat-$\Lambda$CDM model and a $\Lambda$CDM with curvature (denoted as curve-$
    \Lambda$CDM). Through fixing $\Omega_{k}=0$ for flat-$\Lambda$CDM, eq.(\ref{equ:HOM}) can be written as
    \begin{equation}\label{equ:FLCDM}
    H(z)=H_{ 0 }[\Omega _{ m }(1+z)^{ 3 }+ \Omega _{ \Lambda  }]^{ 1/2 }.
    \end{equation}

\subsection{Power-law model}
    Power-law cosmology (denoted as PL) is derived from the power-law relation between the scale factor and proper time as
    \begin{equation}
    \label{equ:PLscale}
    a(t)=\left( \frac{t}{t_0}\right)^{n}.
    \end{equation}
    By using $H \equiv { \dot { a(t) }  } /{ a(t) }$ and $a=(1+z)^{-1}$, one can easily obtain Hubble parameter function as
    \begin{equation}
    \label{equ:PLhubble}
    H(z)=H_{0}(1+z)^ {\frac {1}{n}},
    \end{equation}
    where $H_{0}=n/t_{0}$.

    The $R_{ h}=ct$ can be considered as a unique circumstance of power-law model, in which
    $n=1$. The gravitational horizon scale is handled as $R_{ h}=ct$. Taking $n=1$ into eq.(\ref{equ:PLhubble}),
    we can obtain the Hubble parameter given as
    \begin{equation}
    \label{equ:RHCThubble}
    H(z)=H_{0}(1+z).
    \end{equation}
    The deceleration parameter of $R_{ h}=ct$ model is $q=0$, which expresses a universe expanding steadily. $n>1$ and $n<1$ represent an accelerated and a decelerated universe respectively, according to the relation $q=1/n- 1$.

    After reviewing these two cosmological models, one can derive the comoving distance for each model as below \citep{1999Hogg}.
    The comoving distance for flat-$\Lambda$CDM is written as
    \begin{equation}
    \label{equ:FCMDcomovingdistance}
    D_{ C }=\frac { c }{ H_{0} } \int _{ 0 }^{ z }{ \frac { 1 }{ \sqrt { \Omega _{ m }(1+z)^{ 3 }+\Omega _{ \Lambda  } }  }  } dz.
    \end{equation}
    For curve-$\Lambda$CDM the comoving distance is based on curvature ${ \Omega  }_{ k }$,
    \begin{equation}
    \label{equ:KCMDcomovingdistance}
    D_{C}=\left\{ \begin{array}{ll}
    \frac { c }{ \sqrt {  { \Omega  }_{ k }  }  } \sinh { \left( \int _{ 0 }^{ z }{ \frac { 1 }{ H(z) }  } dz\cdot \sqrt {  { \Omega  }_{ k }   }  \right)  } & { \Omega  }_{ k }>0\\
    \\
    { c }\int _{ 0 }^{ z }{ \frac { 1 }{ H(z) }  } dz\quad & { \Omega  }_{ k }=0 \\
    \\
    \frac { c }{ \sqrt { \left| { \Omega  }_{ k } \right|  }  } \sin { \left( \int _{ 0 }^{ z }{ \frac { 1 }{ H(z) }  } dz\cdot \sqrt { \left|{ \Omega  }_{ k }\right| }  \right)  } & { \Omega  }_{ k }<0
    \end{array} \right . ,
    \end{equation}
    where $H(z)$ is from eq.(\ref{equ:HOM}).
    For power-law model $(n\neq 1)$ we have
    \begin{equation}
    \label{equ:PLcomovingdistance}
    D_{ C }=\frac { c }{ H_{ 0 } } \frac { \left( 1+z \right) ^{ 1-\frac { 1 }{ n }  }-1 }{ 1-\frac { 1 }{ n }  } .
    \end{equation}
    When $n=1$ namely $R_{ h}=ct$ model, the function can be derived as
    \begin{equation}\label{equ:RHCTcomovingdistance}
    D_{ C }=\frac { c }{ H_{ 0 } } \ln \left(1+z\right) .
    \end{equation}
    The angular diameter distance for each model can be written as
    \begin{equation}\label{equ:DaDc}
    D_{ A }=\frac { D_{ C } }{ \left(1+z\right) } .
    \end{equation}
\section{Data and Methodology}\label{sec:samples}
    In this section, we present data including strong gravitational lensing systems, $H(z)$, and BAO. The methods which we treat to various data will be illustrated here.
\subsection{Strong lensing systems}
    Strong gravitational lensing system has become a useful and vital tool in probing cosmology. From the first lensing system has been discovered, there are abundant projects searching for lensing systems, including Sloan Lens ACS (SLACS), BOSS Emission-Line Lens Survey (BELLS), etc. Approximated by singular isothermal sphere (SIS) model or singular isothermal ellipsoid (SIE) model, the Einstein radius can be obtained by measuring the foreground image of Einstein rings and can be expressed through the formula as

\begin{equation}
    \label{equ:SIS}
    \frac{D^{ls}_{A}}{D^{s}_{A}}=\frac{\theta_{ E}\; c^2}{4 \pi \sigma^2_{SIS}}=\mathcal{D}^{obs},
\end{equation}
    where $c$ is the speed of light and $\theta_{ E}$ represents Einstein radius. $D^{ls}_{A}$ and $D^{s}_{A}$ are angular diameter distances from lensing to source and observer to source respectively.

\citet{1996White} found that the dynamical temperature of dark matter halos is larger than measured stellar
velocity dispersion, which hints that the velocity dispersion
$\sigma_{SIS}$ in eq.(\ref{equ:SIS}) may not be equal to
observed stellar velocity dispersion $\sigma_{ap}$.
        \citet{2015Cao} assumed that the mass of strong lensing systems is
        distributed spherically and symmetrically. In their work power-law
        index $\gamma$ of massive elliptical galaxies is treated as a free
        parameter, and $\mathcal{D}^{obs}$ can be derived as
        \begin{equation}
        \label{equ:SIS_cao}
        \mathcal{D}^{obs}=\frac{\theta_{ E}\; c^2}{4 \pi \sigma^2_{ap}} f \left( \theta_{ E}, \theta_{ap},\gamma \right).
        \end{equation}
        $f \left( \theta_{ E}, \theta_{ap}, \gamma \right)$ is a
        complex function related to Einstein radius $\theta_{E}$,
        aperture of certain lensing surveys $\theta_{ap}$, and power-law
        index $\gamma$. But this method may show
        poor sensitivity to cosmological parameters \citep{2016Jie}.

        Another simple method is replacing $\sigma_{SIS}$ of eq.(\ref{equ:SIS})
        by $\sigma_{SIS} = f_e\sigma_{ap}$
        \citep{1992Kochanek,2003Ofek}. As a phenomenological free parameter,
        $f_e$ accounts the systematic errors caused by taking
        $\sigma_{ap}$ as $\sigma_{SIS}$. Unlike \cite{2018Leaf}
        have done recently, free parameter $f_e$ is introduced to calculate
        $\mathcal{D}^{obs}$. Therefore, $\mathcal{D}^{obs}>1$ can not be
        used to exclude unphysical observed lensing systems, which has been
        used in \cite{2018Leaf}.

Moreover, in order to correct observed stellar velocity
dispersion into a circular aperture of radius, $\sigma_0 =
\sigma_{ap}\left(\theta_{eff}/\left(2\theta_{ap}\right)\right)^{-0.04}$
can be used \citep{1995JorgensenA,1995JorgensenB}. $\theta_{eff}$
represents the effective radius, which is achieved by fitting de
Vaucouleurs model \citep{1948Vaucouleurs}. \cite{2015Cao} replaced
$\sigma_{ap}$ of eq.(\ref{equ:SIS_cao}) by $\sigma_0$. Replacing
$\sigma_{SIS}$  directly by $\sigma_{0}$ in eq.(\ref{equ:SIS}) may
not be a proper treatment, which has been used by \cite{2018Leaf}.
Because $\sigma_{0}$ just changes slightly from $\sigma_{ap}$
\citep{2015Cao}.

In order to exclude other systematic errors and unknown
uncertainties introduced from lensing model fitting effect, we use
$\sigma_{ SIS} = f_e\sigma_{ap}$ in this paper, which has been
widely used \citep{2012Cao,2016Liao,2017Li,2017Xia}.

From eq.(\ref{equ:SIS}), we can obtain the angular diameter
distance ratio observed from lensing surveys $\mathcal{D}^{obs}$.
The uncertainty of $\mathcal{D}^{obs}$ can be written as
\begin{equation}\label{equ:ratio uncertainty}
\sigma_{\mathcal{D}^{obs}} =\mathcal{D}^{obs} \sqrt
{\left(\frac{\sigma_{\theta_{E}}}{\theta_{
E}}\right)^{2}+4\left(\frac{\sigma_{ \sigma_{
SIS}}}{\sigma_{SIS}}\right)^{2} }.
\end{equation}

    Following the approach taken by \citet{2008Grillo}, we take uncertainties of Einstein radius as $\sigma_{\theta_{ E}}=0.05\theta_{ E}$.
    The theoretical angular diameter distance ratio $\mathcal{D}^{th}$ can be derived from eqs.(\ref{equ:FCMDcomovingdistance})
    $\sim$ (\ref{equ:DaDc}). The angular diameter distance is written as
    \begin{equation}\label{equ:Dals}
    D_{A}^{ls}=\frac{D_{C}^{ls}}{1+z_{s}}.
    \end{equation}
    Easily,  $\mathcal{D}^{th}$ has the relation
    \begin{equation}\label{equ:ratio theo}
    \mathcal{D}^{th}=\frac{D_{C}^{ls}}{D_{C}^{s}},
    \end{equation}
    where $D_{C}^{ls}=D_{C}^{s}-D_{C}^{l}$ for flat-$\Lambda$CDM,  PL model and $R_{ h}=ct$ model. For curve-$\Lambda$CDM, it is unnecessary to derive an equation relating
    $D_{C}^{ls}$, $D_{C}^{l}$, and $D_{C}^{s}$, but to substitute the range of integration in eq.(\ref{equ:KCMDcomovingdistance}) from $0$ to $z$ into $z_l$ to $z_s$ \citep{2015Syksy}.

    $f_{e}$ as a free parameter in eq.(\ref{equ:SIS}) is required to be fitted simultaneously. Naturally, the mean values
    of parameters can be obtained by minimizing
    \begin{equation}\label{equ:chiSL}
    \chi^{2}_{ SL}=\sum _{ i=1 }^{ n }{ \frac { \left( \mathcal{D}^{th}-\mathcal{D}^{obs} \right) ^{ 2 } }{ \sigma^{2} _{ \mathcal{D}^{obs} } }  } .
    \end{equation}
    The likelihood is $\mathcal{L}_{ SL}\propto \exp(-\chi^{2}_{ SL}/2)$.

    The idea of reducing systematic errors requires a dataset comprehending as much high-quality  data as possible. We utilize 152 strong gravitational lensing systems in our analysis. 118 strong gravitational lensing systems are taken from catalog of \citet{2015Cao}. The catalog was assembled from the Sloan Lens ACS Survey (SLACS), BOSS emission-line lens survey (BELLS), Lens Structure and Dynamics (LSD), and Strong Lensing Legacy Survey (SL2S). Further more, 34 new strong gravitational lensing systems are carefully selected and used in this work. 7 reliable BELLS data with much higher source redshifts are selected cautiously from \citet{2016Shu}. Actually, there are 17 grade-A lenses listed in \citet{2016Shu}. To be specific, five systems require special model-fitting treatments, four systems show significantly larger relative deviations of the Einstein radius, and one system do not meet the selection thresholds. These 10 systems are excluded in order to get rid of some unknown systematic errors imported by any model fitting effect.
    27 SLACS data are chosen from 40 grade-A lenses listed in \citet{2017Shu}. In their work, a parameter $\chi^{2}/dof$ is achieved, which represents the goodness of fitting for SIE model. 27 lenses are filtrated carefully, whose $\chi^{2}/dof$ is within the interval of $1\sigma$ around the constrained value $\chi^{2}/dof=1$. Apart from 118 strong gravitational lensing systems, the new 34 data are listed in Table \ref{newSL}.

\subsection{Hubble parameter H(z)}
By using \textit{cosmic chronometers} approach \citep{2002Jimenez}, the Hubble parameter is measured independently of cosmological models. Through this approach one can directly constrain the expansion history of the universe, and avoid any integrated distance measurements over redshift like SNe Ia and BAOs.

In our work we take 30 $H(z)$ data from \citet{2016Moresco} over 25
    data listed in \citet{2015MeliaA}. Note that all 30 $H(z)$ data are
    cosmological-model independent. The $\chi_{ H(z)}^{2}$ for Hubble
    parameters is given by
    \begin{equation}\label{equ:chiHZ}
    \chi^{2}_{ H(z)}=\sum _{ i=1 }^{ n }{ \frac { \left( {H(z)}^{th}-{H(z)}^{obs} \right) ^{ 2 } }{ \sigma ^{2}_{ {H(z)}^{obs} } }  } ,
    \end{equation}
    where $H(z)^{th}$ is the theoretical value of Hubble parameter. The likelihood is $\mathcal{L}_{H(z)}\propto \exp(-\chi^{2}_{ H(z)}/2)$.

\subsection{Baryon acoustic oscillations}\label{sec:BAO}
As a model-independent standard ruler, BAO is a practical tool in
probing cosmology. Through measuring the comoving sound horizon of
the baryon drag epoch at different redshifts \citep{2005Eisenstein},
the sound horizon size at the end of the drag era $z_d$ is written
as
    \begin{equation}\label{equ:rd}
    { r }_{ d }=r_{s}(z_{d})=\int _{ { z }_{ d } }^{ \infty  }{ \frac { { c }_{  s }(z)dz }{ H(z) }  } ,
    \end{equation}
where $c_{s}$ is the speed of sound. $r_d$ may relate more to early
universe, from which we can't get enough information. Besides
$\Lambda$CDM, other cosmological models may not have a compatible
function of $r_d$ \citep{2017Verdea}. Considering our work focus on
different cosmological models, we regard $r_d$ as a free parameter
following. BAO measures the ratio of comoving sound horizon and the
distance scale \citep{2005Eisenstein} given as
    \begin{equation}\label{equ:BAOratio}
    R(z)=\frac{r_{d}}{D_{V}},
    \end{equation}
    and $D_{V}$ is defined as
    \begin{equation}\label{equ:DV}
    { D }_{ V }(z)\equiv \left[ D_{ C }^{ 2 }(z)\frac { cz }{ H(z) }  \right]^{\frac{1}{3}},
    \end{equation}
    where comoving distance $D_{C}$ and Hubble parameter $H(z)$ are both defined in Section \ref{sec:Model}.

We use \textbf{11} BAO measurements from different surveys, which are
listed in Table \ref{tab:BAO}. In order to avoid importing BAO data
at same redshifts repeatedly, we use the latest measurements of each
redshift. Also we avoid importing BAOs with correlated surveys as far as possible \citep{2018Yu}.
To be specific, BAOs from 6dFGS, SDSS DR7, BOSS DR12, BOSS DR14 and Ly-$\alpha$ are
uncorrelated with each other. Actually, there are two methods of Ly$\alpha$-forest. One method is Ly$\alpha$ auto-correlation, and the other is Ly$\alpha$ cross-correlation. Here, we use the
 BAO of combining both these two methods from \cite{2017Mas}, who have combined
 their measurement by cross-correlation at redshift 2.4 and auto-correlation at redshift 2.33 \citep{2017Bautista} together.

Basically, there are two main analytical methods of
measuring BAOs. One is anisotropic galaxy clustering measurement
which constrains comoving distance $D_{M}(z)$ in the transverse direction
 and Hubble parameter $H(z)$ along line-of-sight. In this work,
  $D_M$ equals to $D_C$ which has been defined for different cosmological
   models in section \ref{sec:Model}.
This method is also known as an application of Alcock-Paczynski
effect \citep{1979Alcock}. The other is spherically averaged
clustering which constrains volume-averaged distance $D_V$
\citep{2005Eisenstein}.

Results of both measurements are all used in this work.
The original data of each BAO are listed in Table \ref{tab:BAO}.
Specially, for
different BAO measurements at $z = 0.38, 0.51, 0.61$, measurements in the transverse direction and along line-of-sight are both included \citep[following e.g.][]{2017Wojtak, 2018Lemos, 2018Ryan}. So we take covariance matrix from
  \citet{2017Alam} into our calculation. For BAO at $z=2.4$ \citep{2017Mas}, we also take a small covariance matrix.

The $\chi^{2}_{BAO}$ is given by
\begin{equation}\label{equ:chiBAO}
\chi^{2}_{ BAO}=  ({\bm{V}^{obs}-\bm{V}^{th}})\bm{C}^{-1}(\bm {V}^{obs}-\bm{V}^{th})^{T},
\end{equation}
where ${\bm{V}^{obs}}$ and ${\bm{V}^{th}}$ are observed measurements
and theoretical values corresponding to second column of Table \ref{tab:BAO}.
$\bm{C}^{-1}$ is inverse covariance matrix of the observed variables. For BAOs at $z = 0.38, 0.51, 0.61$, the corresponding elements of inverse covariance matrix in eq.(\ref{equ:chiBAO}) are:
$$\begin{array}{lll}
c_{4 4}=2.88, & c_{4 5}=-1.24,& c_{4 6}=0.16,\\
c_{4 7}=-6.76, & c_{4 8}=3.61,& c_{4 9}= -0.58,\\
c_{5 5}=2.49, & c_{5 6}=-0.90,& c_{5 7}=2.01,\\
c_{5 8}=-7.14, & c_{5 9}=3.21,& c_{6 6}=1.40,\\
c_{6 7}=-0.66, & c_{6 8}=2.28,& c_{6 9}= -4.82,\\
c_{7 7}=380.85, & c_{7 8}=-192.93,& c_{7 9}= 18.60,\\
c_{8 8}=510.88,& c_{8 9}= -210.83,& c_{9 9}= 339.81.
\end{array}$$
For BAO at $z=2.4$, the inverse covariance matrix is:
$$\begin{array}{lll}
c_{10 10}=799.17, & c_{10 11}=1578.
\end{array}$$
These elements should be multiplied by $10^{-3}$. Here, $c_{ij}=c_{ji} \; \left( i\neq j \right)$. For the rest of elements, we have $c_{ii}=1/\sigma_{i}^{2}$ and $c_{ij}=c_{ji}=0 \; \left( i\neq j \right)$, where $\sigma_{i}^{2}$ is error for BAO measurements.

The likelihood is $\mathcal{L}_{BAO}\propto \exp(-\chi^{2}_{BAO}/2)$.
\\

In our work, we constrain the parameters for each cosmological model
by using different combinations of these three kinds of data. The
free parameters $f_e$ and $r_d$ are also needed to be constrained
simultaneously. The three different datasets and corresponding total
likelihoods are listed as below.
    \begin{itemize}
        \item Strong gravitational lensing system only (hereafter SL): $\mathcal{L}_{ SL}$
        \item Strong gravitational lensing system and $H(z)$ (hereafter SL+$ H(z)$): $\mathcal{L}_{ SL}\cdot \mathcal{L}_{ H(z)}$
        \item Strong gravitational lensing system, $ H(z)$ and BAO (hereafter SL+$ H(z)$+BAO) $\mathcal{L}_{ SL}\cdot \mathcal{L}_{ H(z)}\cdot \mathcal{L}_{ BAO}$
    \end{itemize}

It should be noted that the maximum likelihood method can be achieved by
Bayesian approach \citep[e.g.][]{2005DAgostini,2010Hogg} or
frequentist approach \citep[e.g.][]{2016bPlanck}. In this work, we use maximum likelihood method basing on Bayesian approach.
    Using information criteria may not be an excellent choice,
    because the information criteria derived from Bayesian approach
    may be a little different from
     the true values of comparison criteria. Also, the information criteria do not include the
     prior probability distributions into comparison \citep{2007Liddel}. So, we then introduce Bayesian
      evidence \citep{2007Liddel,2008Trotta} as model comparison criteria. The Bayesian evidence is defined as
\begin{equation}
\label{equ:BE}
E\equiv \int { \mathcal{L}(\theta ) } P(\theta) \rm d\theta,
\end{equation}
where $\theta$ represents parameters of models. $\mathcal{L}(\theta)$ is likelihood function,
and $P(\theta)$ gives priors distribution of parameters. The Bayes factor in natural logarithm $\ln B$ can
be derived as
\begin{equation}
\label{equ:Bfactor}
\ln B = \ln E_{fid} - \ln E_{\alpha},
\end{equation}
where $fid$ represents flat-$\Lambda$CDM which we fix as the fiducial model, and ${\alpha}$ represent other models in one-on-one comparison.
The $\ln E$ represents Bayesian evidence (eq.\ref{equ:BE}) in natural logarithm.
The preference strength for flat-$\Lambda$CDM  can be described as weak, moderate or strong,
according to $\ln B > 1.0,$ $2.5$ or $5.0$ \citep{2008Trotta}.
Naturally, a negative $\ln B$ value represents flat-$\Lambda$CDM is not favored.

\section{Results}\label{sec:result}
We use an open source python package {emcee} \citep{2013Foreman} to
constrain cosmological parameters through Markov chain Monte Carlo
method. The python package {nestle}\footnote{https://github.com/kbarbary/nestle}
is used to calculate Bayesian evidence through nested sampling algorithm \citep{2004Skilling}.
The Hubble constant $H_{0}$ and the comoving sound horizon
$r_{d}$ are replaced by $h_0=H_{0}/100$ and $r_{*}=r_{d}/100$ in our
numeration. Uniform distributions of prior probabilities for the
parameters are assumed: $P(\Omega_{ m })=U[0,1]$, $P(\Omega_{ k
})=U[-1,1]$, $P(n)=U[0,2]$, $P(f_{e})=U[0.5,1.5]$, and
$P(h_{0})=U[0.5,1]$. When we constrain
parameters of curve-$\Lambda$CDM, we add another prior $P(\Omega_{
    \Lambda })=U[0.6,0.8]$ to constrain $\Omega_{ k }$ more precisely.

\cite{2017Verdeb} constrained the early cosmology from
current CMB observations without any assumptions of late-time
cosmology. They model-independently measured the value of $r_d$ as $r_d=147.4\pm 0.7$ Mpc.
Because BAOs are sensitive to $r_{d}$ and $H_{0}$, BAOs may not be able to constrain both $r_{d}$ and $H_{0}$ at same time. So,
we set uniform prior of $r_{*}$ as $P(r_{*})=U[1,2]$ or Gaussian prios $r_{*} = 1.474 \pm 0.007$ when importing BAO measurements, respectively.
Meanwhile, we set $\Omega_{ \Lambda  }$ as a free parameter in curve-$\Lambda$CDM by only setting uniform priors on $\Omega_{ k }$ and $\Omega_{ m }$.

Due to the poor quality of SL data, it may not properly constrain cosmological parameters. In Table \ref{tab:parameters}, we only list the mean value and 1$\sigma$ limits of each parameter. According to Bayes factors listed in Table
\ref{tab:IC}, flat-$\Lambda$CDM is
slightly favored rather than $R_h=ct$, according to $\ln B > 2.5$.
However, \citet{2018Leaf} found $R_{ h}=ct$ model is preferred over
$\Lambda$CDM model. The reason is that they excluded some SL
systems, as we discuss in section \ref{sec:samples}.

Then another dataset SL+$ H(z)$ is imported in our analysis. The
fitting results are listed in the second four rows of Table
\ref{tab:parameters}. By comparing the result only fitted by SL, one
can easily find that the accuracies of $\Omega_{ m }$ and $n$ are
both improved dramatically. There is no significant variation of the
free parameter $f_e$ for different models. Till now, we believe that
through combination of different datasets, the accuracy of fitting
will be enhanced significantly. By comparing Bayes factors in Table
\ref{tab:IC}, it is impossible to rule out some models from
flat-$\Lambda$CDM. Only PL can be
ruled out, according to $\ln B > 2.5$.

\begin{figure*}

        \includegraphics[width=0.45\linewidth]{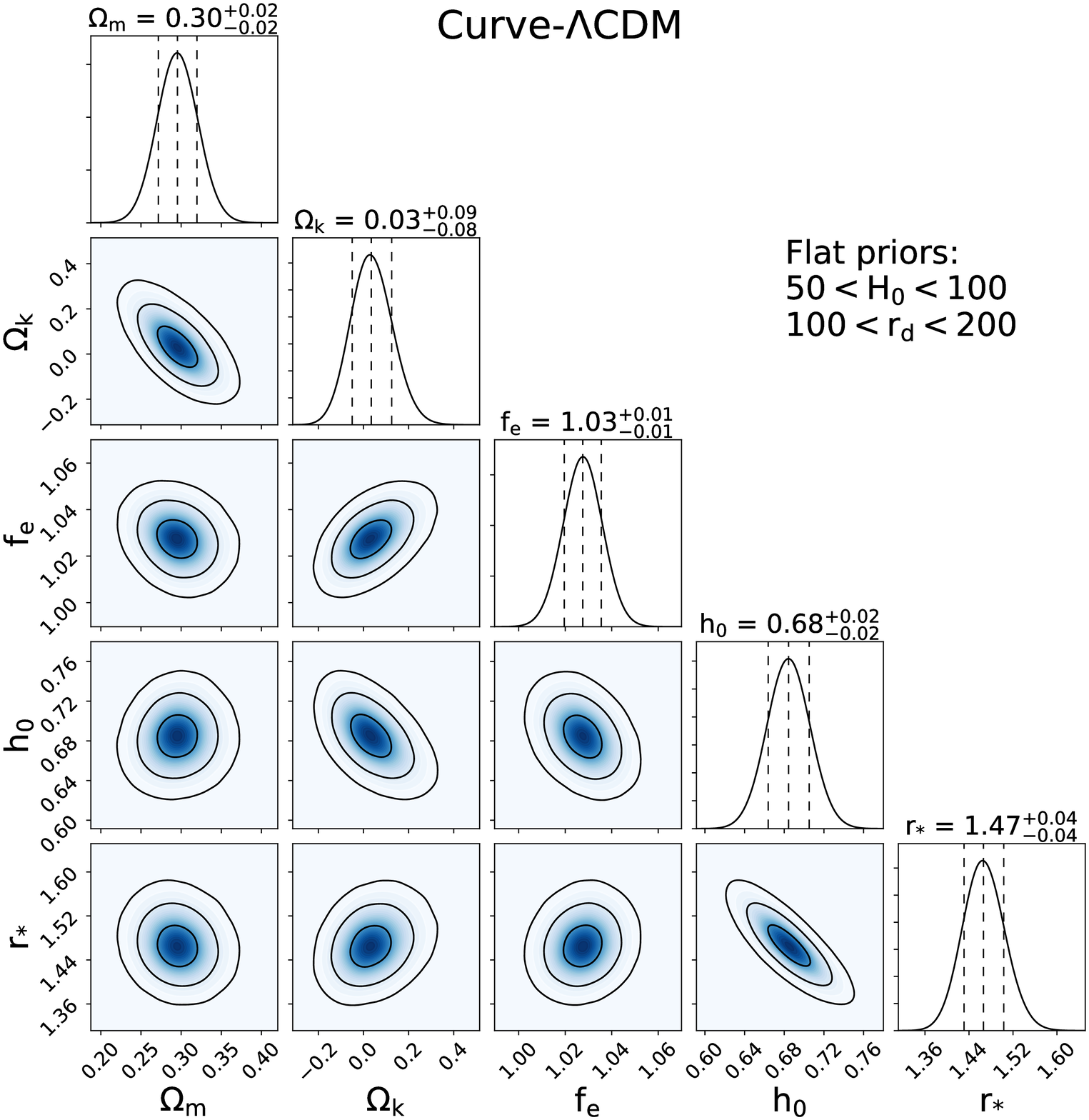}\hspace{1cm}
        \includegraphics[width=0.45\linewidth]{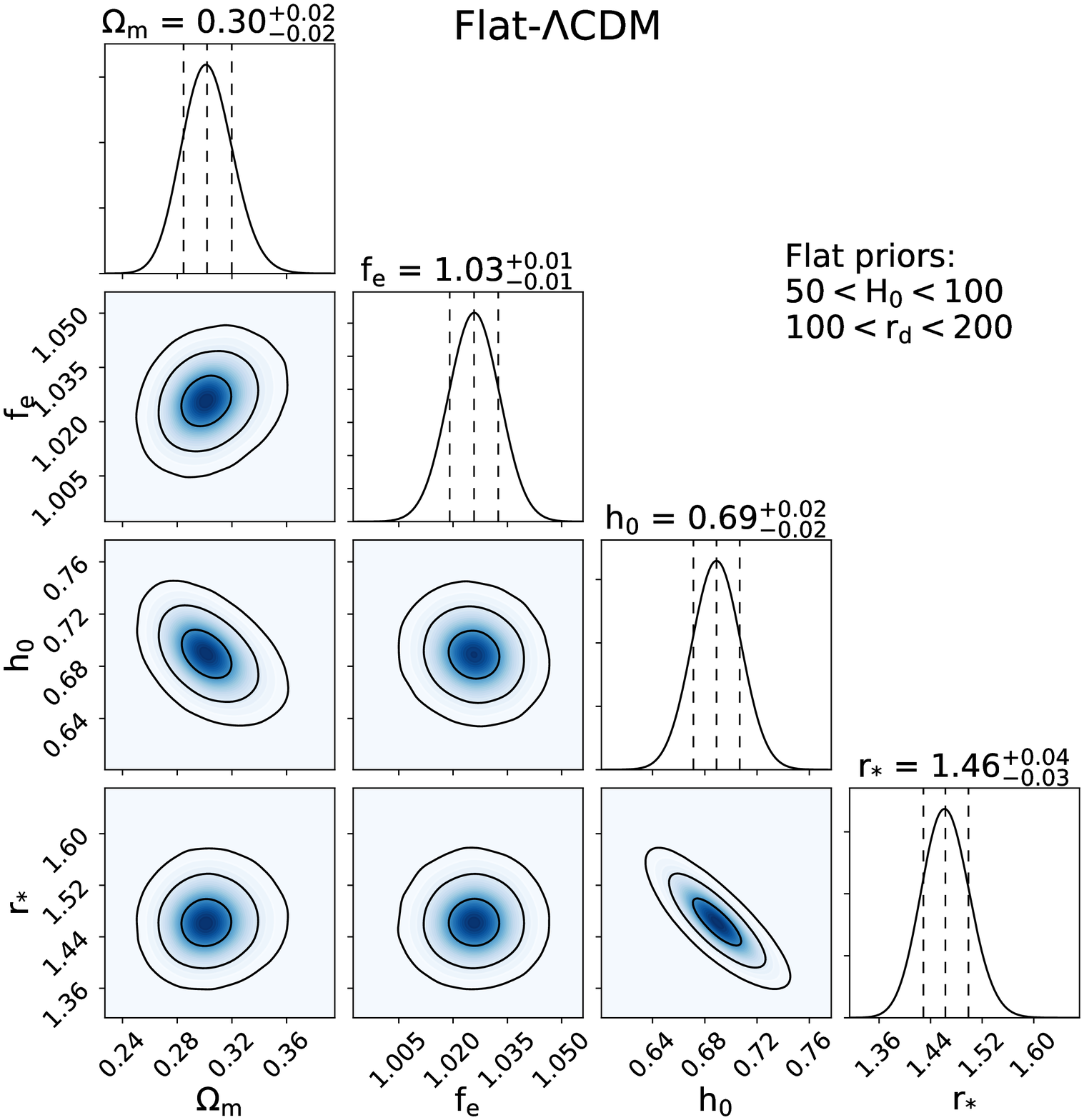} \\ \vspace{1cm}
        \includegraphics[width=0.45\linewidth]{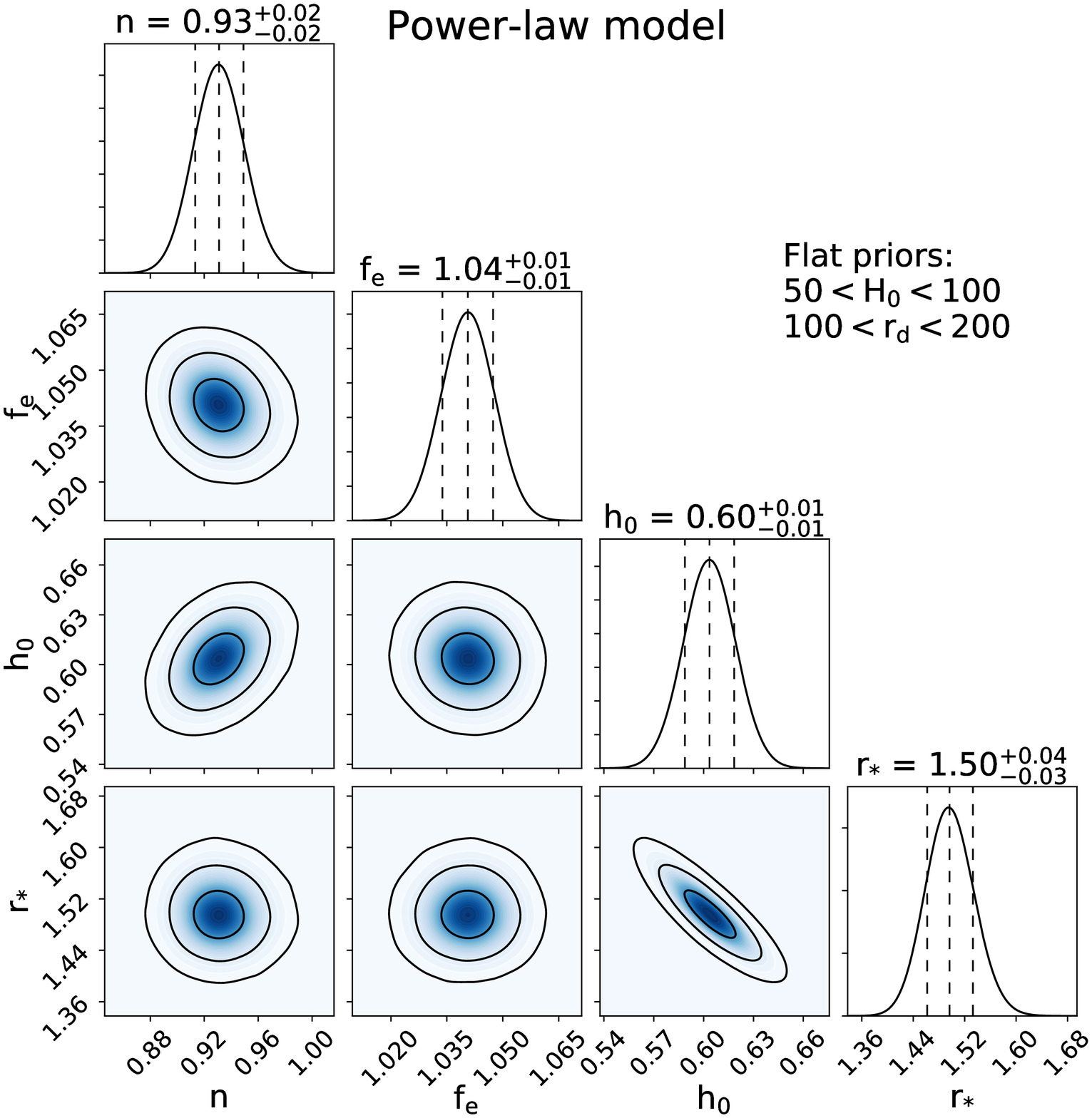} \hspace{1cm}
        \includegraphics[width=0.45\linewidth]{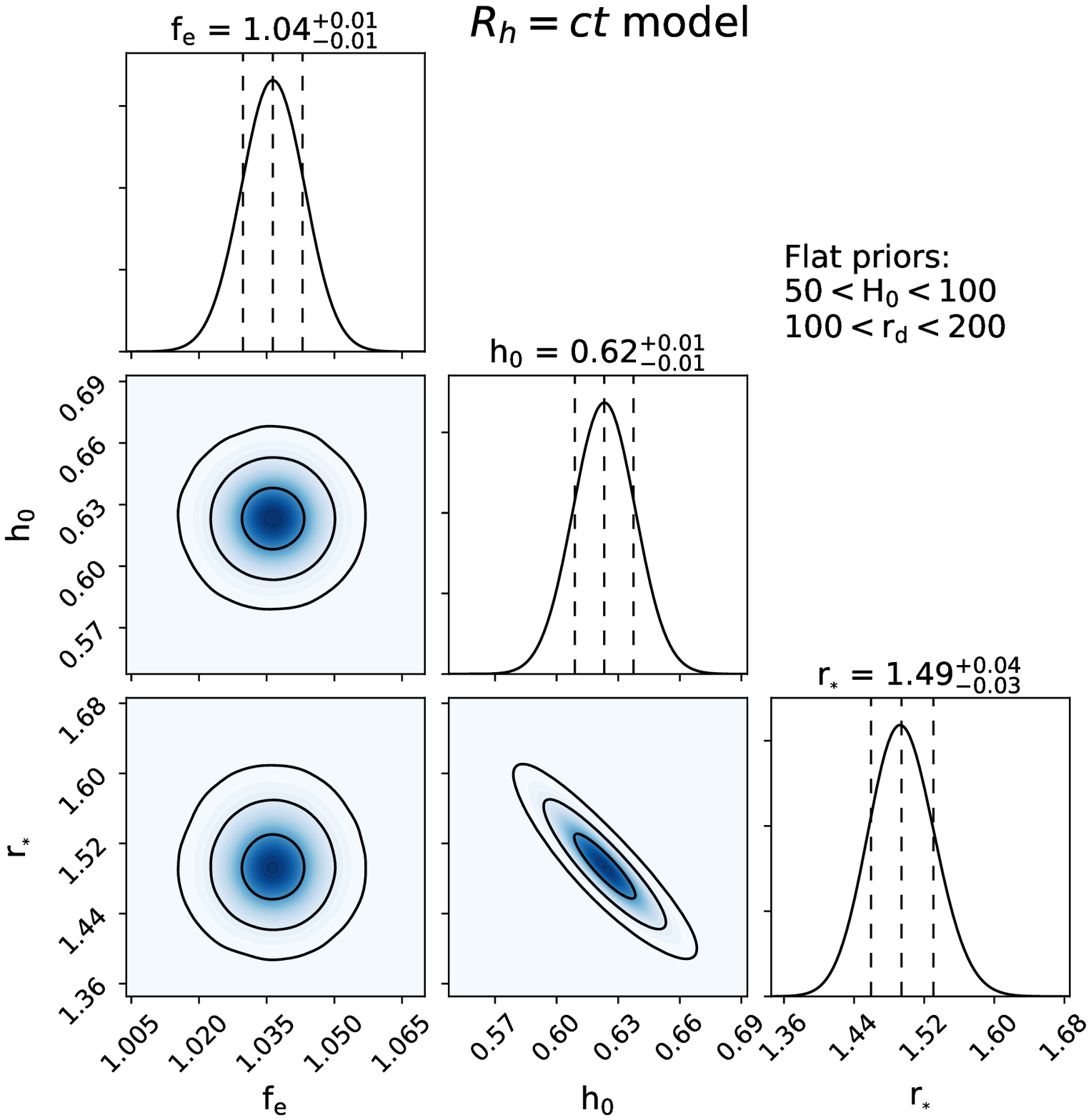}
    \caption{Fitting results of different models from
        SL+$ H(z)$+BAO for a flat prior on $r_{d}$ ($P(r_{*})=U[1,2]$). The three circles represent 1$\sigma$, 2$\sigma$ and
        3$\sigma$ uncertainties. Vertical lines given in diagonal hist
        graphics are 1$\sigma$ interval for each parameter.}
    \label{fig:prior_rd}
\end{figure*}

Next, the last dataset SL+$ H(z)$+BAO is applied to our work.
Firstly, we use a flat prior of $r_{d}$ , and our calculation
    gives optimized fitting results, which are listed in table \ref{tab:parameters} (called case 1). To be specific,
    the top-left panel of Figure \ref{fig:prior_rd} reveals the results of MCMC simulation
for curve-$\Lambda$CDM. The mean values are $\Omega_{ m
}=0.30\pm 0.02$, $\Omega_{ k }=0.03\pm 0.09$,
$h_{0}= 0.68\pm 0.02$, and $r_{*}=1.47\pm 0.04$. The value of
$\Omega_{ k }$ is consistent with zero at $1\sigma$ confidence
interval. The fitting results of flat-$\Lambda$CDM are shown in top-right panel of Figure
\ref{fig:prior_rd}, where $\Omega_{ m }=0.30\pm 0.02$, $h_{0}= 0.69\pm 0.02$, and $r_{*}=1.46\pm
0.04$. For these two models, $r_{*}$ and $h_{0}$ are
consistent. In the bottom-left panel of Figure \ref{fig:prior_rd}, constraints on PL model are
$n=0.93\pm 0.02$, $h_{0}= 0.60\pm 0.01$, and
$r_{*}=1.50\pm0.04$. For $R_{h}=ct$ model, the bottom-right panel of Figure
\ref{fig:prior_rd} reveals the mean values as
$h_{0}= 0.62\pm 0.01$, and $r_{*}=1.49\pm0.04$.
Values of $r_{d}=100r_{*}$ for these two models are a little bit larger than $r_d$ measurement as
$r_{d}=147.4\pm 0.7 {\rm Mpc}$ \citep{2017Verdeb}.
 Table \ref{tab:IC} shows a
 large improvement of fitting by adding BAO data. It is possible to
rule out some models. We get the Bayes factors of flat-$\Lambda$CDM
versus PL model is 7.33, and versus $R_{ h}=ct$ is 10.18. So the dataset prefers
the flat-$\Lambda$CDM.
But Bayes factor is not large enough to compare flat-$\Lambda$CDM and
 curve-$\Lambda$CDM.

\begin{figure*}

    \includegraphics[width=0.45\linewidth]{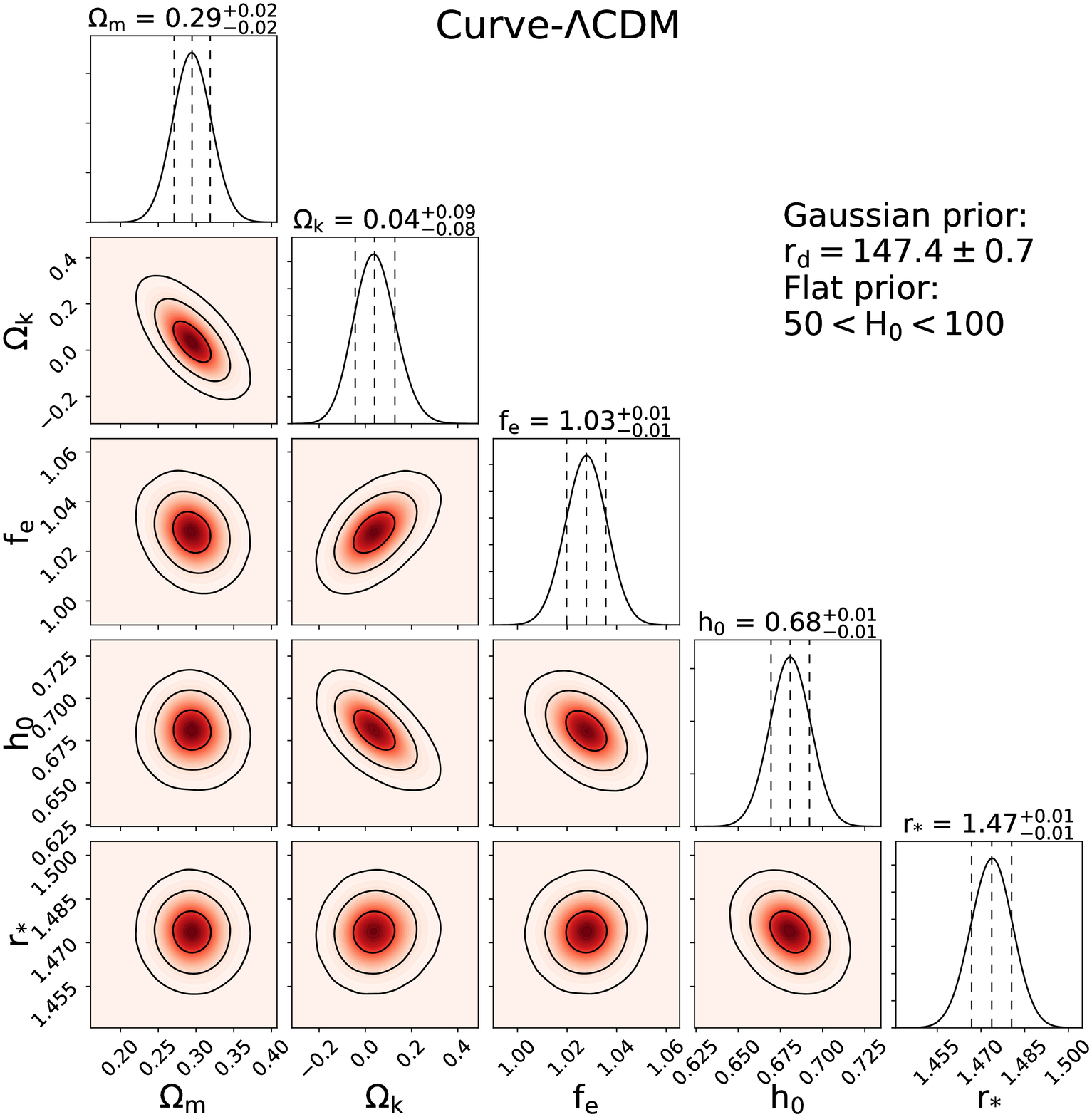}\hspace{1cm}
    \includegraphics[width=0.45\linewidth]{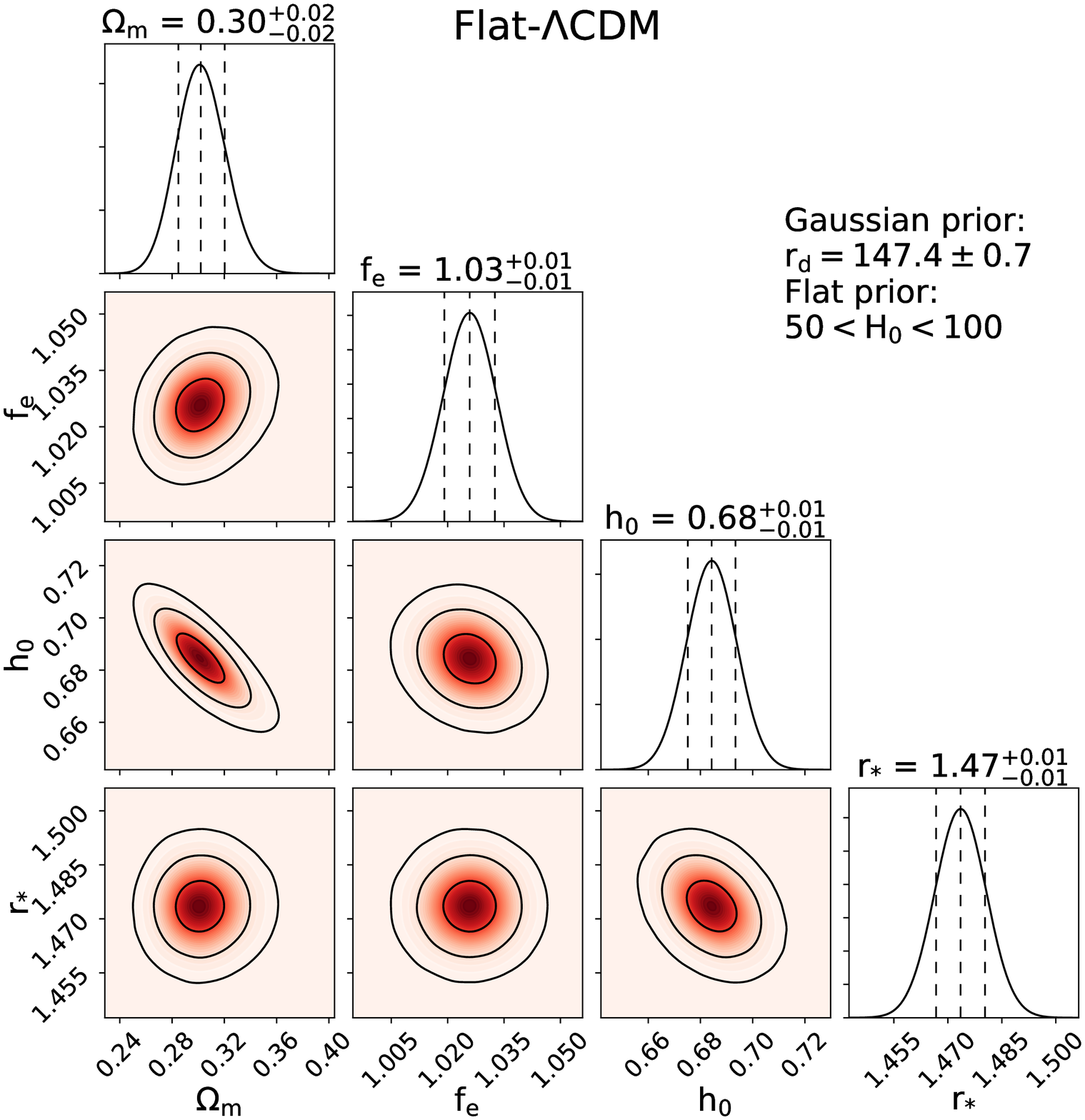} \\
    \vspace{1cm}
    \includegraphics[width=0.45\linewidth]{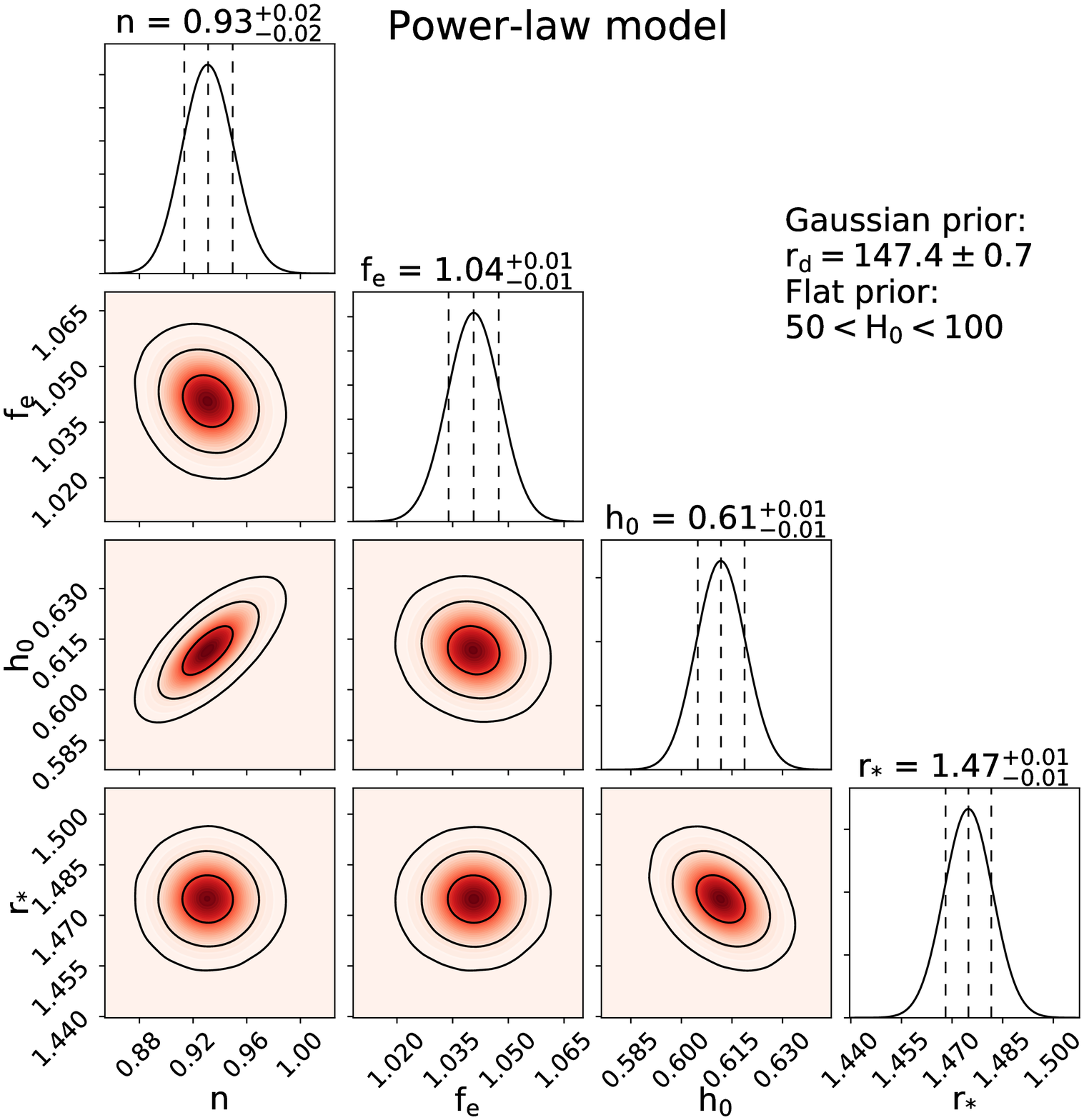} \hspace{1cm}
    \includegraphics[width=0.45\linewidth]{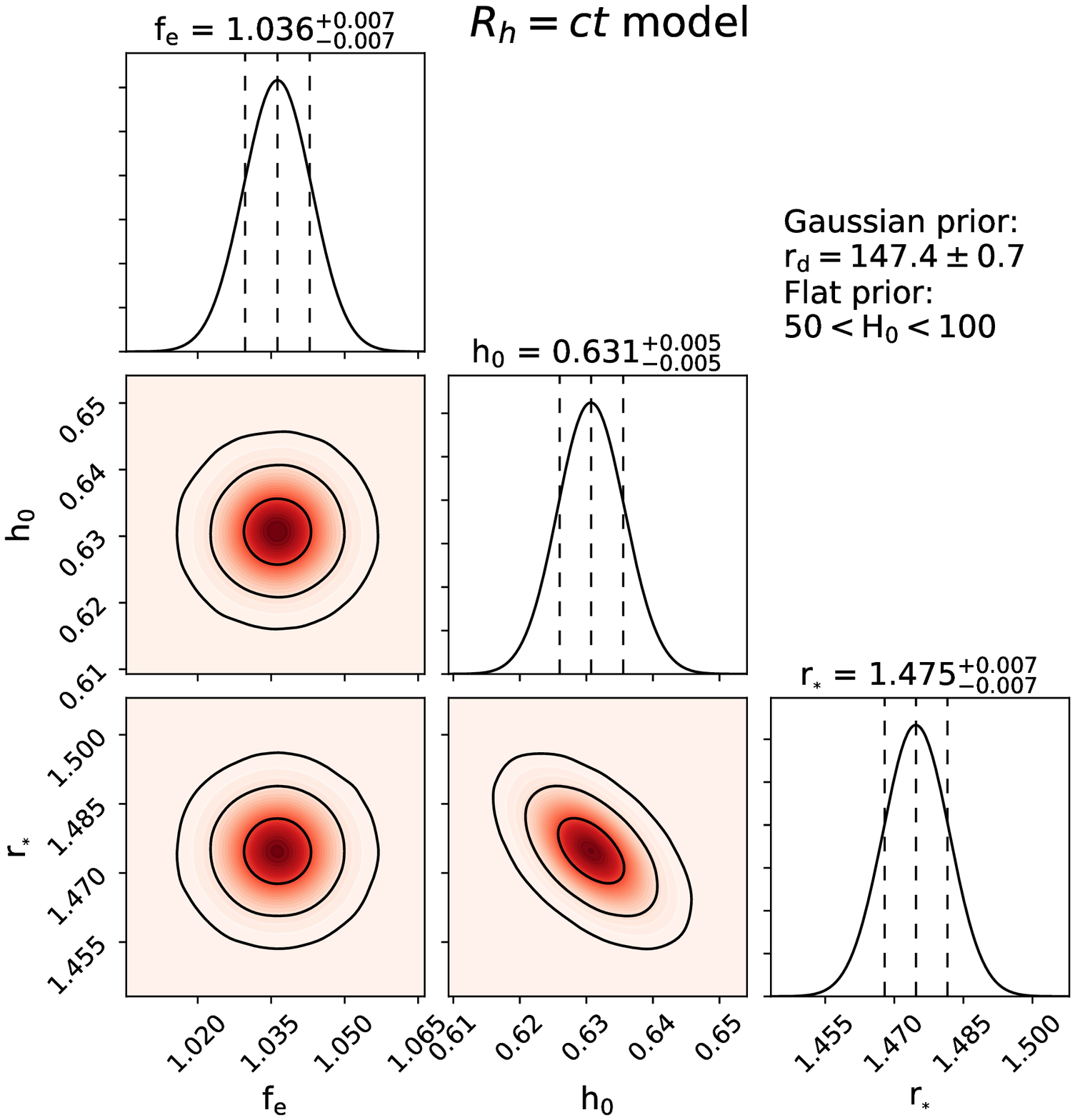}
    \caption{Fitting results of different models from SL+$ H(z)$+BAO under the Gaussian prior of $r_d$ as $r_{*}=1.474\pm 0.007$.}
    \label{fig:gau_rd}
\end{figure*}

We then only set a Gaussian prior $r_{*} = 1.474 \pm 0.007$ to apply SL+$ H(z)$+BAO. In contrast to
flat prior of $r_{*}$,
we denote the results as case 2 in Tables \ref{tab:parameters} and \ref{tab:IC}.
We also set red density contours in Figure \ref{fig:gau_rd}.
The mean value of $\Omega_{ m }$ and $\Omega_{ k }$ for flat-$\Lambda$CDM and
curve-$\Lambda$CDM are slightly changed.
For flat-$\Lambda$CDM, $\Omega_{ m} = 0.30\pm 0.02$.
For curve-$\Lambda$CDM, $\Omega_{ m} = 0.29\pm 0.02$ and $\Omega_{ k }=0.04\pm 0.09$.
And the values of $r_{d} = 100r_{*}$ for these four models basically equal to  $r_{d}=147.4\pm 0.7\, {\rm Mpc}$ \citep{2017Verdeb}. In Figure \ref{fig:gau_rd}, as what we expected, setting Gaussian prior of $r_d$ breaks the degeneracy between $r_{d}$ and $H_{0}$. It slightly increases the value of Hubble parameter for power-law and $R_{h}=ct$ models from Table \ref{tab:parameters}. But the model preference is not enhanced significantly from case 2 in Table \ref{tab:IC}.
According to Bayes factors in Table \ref{tab:IC}, flat-$\Lambda$CDM is still the data-favorable model rather than PL and $R_h=ct$ models.

One may see the degeneracy between $r_{d}$ and $H_{0}$ in Figure \ref{fig:prior_rd}. If we only set a large Gaussian prior on $H_{0}$, the $r_{d}$ will be reduced significantly. We have certified this idea. Because the value of $r_{d}$ is not properly constrained, we do not list the result here. As shown in Table \ref{tab:parameters}, the case 2 also shows lower values of $H_{0}$ for power-law and $R_{h}=ct$ models than $\Lambda$CDM model. So, we set Gaussian priors not only on $r_{d}$, but also on $H_{0}$. The main reason for setting these two Gaussian priors is that we try to properly constrain $r_{*}$ and $h_{0}$ for power-law and $R_{h}=ct$ models, and to improve the preferences of models. The results are listed in Table \ref{tab:parameters}, and Bayes factors in Table \ref{tab:IC}. Except the same Gaussian prior of $r_{d}$ is set both for case 3 and case 4, case 3 represents the Gaussian prior on $H_{0}$ as $H_{0} = 73.52 \pm 1.62\, \rm{km\,s^{-1}\,Mpc^{-1}}$ from \citet{2018Riess}, and case 4 represents the Gaussian prior $H_{0} = 67.80 \pm 0.09\,\rm{km\,s^{-1}\,Mpc^{-1}}$ from \citet{2016Planck}. In Table \ref{tab:IC}, we can see that the model preferences are increased. The power-law and $R_{h}=ct$ models are penalized more by the additional Gaussian priors on $H_{0}$. However, it can not give a strong evidence to support flat-$\Lambda$CDM over curve-$\Lambda$CDM.

\section{Conclusions}\label{sec:discuss}
In this work, we use 152 strong gravitational lensing systems, 30
    $ H(z)$ and \textbf{11} BAO data to compare cosmological models. In all cases, we find that the flat-$\Lambda$CDM is strongly preferred over the power-law and $R_{h}=ct$ models.
    The corresponding probability of data-favorable model at $\ln B >5$ is greater than $99.3\%$ \citep{2008Trotta},
     The flat-$\Lambda$CDM is not obviously preferred over the
    curve-$\Lambda$CDM according to that the Bayes factors
    give weak evidence to distinguish them.
    The value of $n$ for power-law cosmology become close to $1$ as more datasets are
    considered.

    By using strong gravitational lensing systems only, the capability
    of discriminating cosmological models may not be reliable. This may due to the
    absence of accuracy in measuring Einstein radius and stellar
    velocity dispersion. We roughly give a correction $f_e$ to all
    lensing systems from different surveys, as what has been done
    \citep{2012Cao,2016Liao}. It is important to consider an
    effective method targeting at correction of velocity dispersion
    like the work done by \citet{2015Cao}. This task should
    consider of systematic errors imported from different lensing
    surveys. Strong gravitational lensing systems are equipped with
    intrinsic goodness. We still look forward to probe cosmological models
    using strong gravitational lensing systems only. Because of the
    limitation of lacking enormous database, combining
    other kinds of data seems like an efficient and
    straightforward way to probe cosmic acceleration.

\section*{Acknowledgements}
We thank an anonymous referee for useful suggestions and comments.
We would like to thank G.Q. Zhang, Z.Q. Sun, Y.Y. Wang, and Hai
Yu for helpful discussion. This work is supported by the National Natural Science Foundation of China
(grant U1831207).

        \begin{table*}
        \begin{center}
            \renewcommand{\arraystretch}{1.5}
            \addtolength{\tabcolsep}{+10pt}
            \caption{Compilation of New Strong Lensing Systems.}
            \label{newSL}
            \begin{tabular}{@{}cccccc@{}}
                \hline
                \hline
                Name       & $z_{l}$ & $z_{s}$ & $\sigma_{ ap} ( km\;  s^{-1})$ & $\theta_{ E}( ^{\prime \prime} )$ & Survey \\ \hline
                J1110$+$2808 & 0.6073  & 2.3999  & 191$\pm$39                & 0.98                         & BELLS  \\
                J2342$-$0120 & 0.5270   & 2.2649  & 274$\pm$43                & 1.11                         & BELLS  \\
                J1110$+$3649 & 0.7330   & 2.5024  & 531$\pm$165               & 1.16                         & BELLS  \\
                J1201$+$4743 & 0.5628  & 2.1258  & 239$\pm$43                & 1.18                         & BELLS  \\
                J0742$+$3341 & 0.4936  & 2.3633  & 218$\pm$28                & 1.22                         & BELLS  \\
                J1141$+$2216 & 0.5858  & 2.7624  & 285$\pm$44                & 1.27                         & BELLS  \\
                J0029$+$2544 & 0.5869  & 2.4504  & 241$\pm$45                & 1.34                         & BELLS  \\
                J0159$-$0006 & 0.1584  & 0.7477  & 216$\pm$18                & 0.92                         & SLACS  \\
                J1330$+$1750 & 0.2074  & 0.3717  & 250$\pm$12                & 1.01                         & SLACS  \\
                J1301$+$0834 & 0.0902  & 0.5331  & 178$\pm$8                 & 1.00                            & SLACS  \\
                J1010$+$3124 & 0.1668  & 0.4245  & 221$\pm$11                & 1.14                         & SLACS  \\
                J1048$+$1313 & 0.1330   & 0.6679  & 195$\pm$10                & 1.18                         & SLACS  \\
                J1550$+$2020 & 0.1351  & 0.3501  & 243$\pm$9                 & 1.01                         & SLACS  \\
                J1430$+$6104 & 0.1688  & 0.6537  & 180$\pm$15                & 1.00                            & SLACS  \\
                J0955$+$3014 & 0.3214  & 0.4671  & 271$\pm$33                & 0.54                         & SLACS  \\
                J0324$-$0110 & 0.4456  & 0.6239  & 310$\pm$38                & 0.63                         & SLACS  \\
                J1041$+$0112 & 0.1006  & 0.2172  & 200$\pm$7                 & 0.60                          & SLACS  \\
                J1541$+$3642 & 0.1406  & 0.7389  & 194$\pm$11                & 1.17                         & SLACS  \\
                J1127$+$2312 & 0.1303  & 0.3610   & 230$\pm$9                 & 1.25                         & SLACS  \\
                J1137$+$1818 & 0.1241  & 0.4627  & 222$\pm$8                 & 1.29                         & SLACS  \\
                J1051$+$4439 & 0.1634  & 0.5380   & 216$\pm$16                & 0.99                         & SLACS  \\
                J1553$+$3004 & 0.1604  & 0.5663  & 194$\pm$15                & 0.84                         & SLACS  \\
                J1101$+$1523 & 0.1780   & 0.5169  & 270$\pm$15                & 1.18                         & SLACS  \\
                J0920$+$3028 & 0.2881  & 0.3918  & 297$\pm$17                & 0.70                          & SLACS  \\
                J0754$+$1927 & 0.1534  & 0.7401  & 193$\pm$16                & 1.04                         & SLACS  \\
                J1607$+$2147 & 0.2089  & 0.4865  & 197$\pm$16                & 0.57                         & SLACS  \\
                J0757$+$1956 & 0.1206  & 0.8326  & 206$\pm$11                & 1.62                         & SLACS  \\
                J1056$+$4141 & 0.1343  & 0.8318  & 157$\pm$10                & 0.72                         & SLACS  \\
                J1142$+$2509 & 0.1640   & 0.6595  & 159$\pm$10                & 0.79                         & SLACS  \\
                J0143$-$1006 & 0.2210   & 1.1046  & 203$\pm$17                & 1.23                         & SLACS  \\
                J0851$+$0505 & 0.1276  & 0.6371  & 175$\pm$11                & 0.91                         & SLACS  \\
                J0847$+$2348 & 0.1551  & 0.5327  & 199$\pm$16                & 0.96                         & SLACS  \\
                J0956$+$5539 & 0.1959  & 0.8483  & 188$\pm$11                & 1.17                         & SLACS  \\
                J1144$+$0436 & 0.1036  & 0.2551  & 207$\pm$14                & 0.76                         & SLACS  \\ \hline
            \end{tabular}

        \end{center}
        \begin{flushleft}
            \textsc{Note.}\\
            Column 1 is the name of strong gravitational
            lensing system. Columns 2 and 3 are redshifts for lensing and source respectively.
            Column 4 is aperture velocity dispersion observed from spectrum which we take as stellar velocity dispersion $\sigma_{ ap}$.
            Column 5 and 6 are the Einstein radius and name of the surveys, respectively. \\
            \vspace{1em}
        \end{flushleft}
    \end{table*}

\begin{table*}
    \renewcommand{\arraystretch}{1.5}
    \addtolength{\tabcolsep}{+2pt}
    \centering
    \caption{BAOs from different surveys.}
    \label{tab:BAO}

\begin{tabular}{clcccc}
    \hline
    Redshift & Measurement              & Value             & $r_{\rm fid}$ & Survey    & Refs                \\ \hline
    0.106    & $r_{d}/D_{V}$            & $0.336 \pm 0.015$ & ---           & 6dFGS     & \citet{2011Beutler} \\
    0.15     & $D_{V}r_{\rm fid}/r_{d}$ & $664 \pm 25 $     & 148.69        & SDSS DR7  & \citet{2015Ross}    \\
    1.52     & $D_{V}r_{\rm fid}/r_{d}$ & $3843 \pm 147 $   & 147.78        & SDSS DR14 & \citet{2018Ata}     \\
    0.38     & $D_{M}r_{\rm fid}/r_{d}$ & $1518 \pm 22 $    & 147.78        & SDSS DR12 & \citet{2017Alam}    \\
    0.51     & $D_{M}r_{\rm fid}/r_{d}$ & $1977 \pm 27 $     & 147.78        & SDSS DR12 & \citet{2017Alam}    \\
    0.61     & $D_{M}r_{\rm fid}/r_{d}$ & $2283 \pm 32 $    & 147.78        & SDSS DR12 & \citet{2017Alam}    \\
    0.38     & $H(z) r_{d}/r_{\rm fid}$    & $81.5 \pm 1.9 $   & 147.78        & SDSS DR12 & \citet{2017Alam}    \\
    0.51     & $H(z) r_{d}/r_{\rm fid}$    & $90.4 \pm 1.9 $   & 147.78        & SDSS DR12 & \citet{2017Alam}    \\
    0.61     & $H(z) r_{d}/r_{\rm fid}$    & $97.3 \pm 2.1 $   & 147.78        & SDSS DR12 & \citet{2017Alam}    \\
    2.40     & $D_{M}/r_{d}$            & $36.6 \pm 1.2$  & ---           & SDSS DR12 & \citet{2017Mas}     \\
    2.40     & $c/H(z) r_{d}$              & $8.94 \pm 0.22$   & ---           & SDSS DR12 & \citet{2017Mas}     \\ \hline
\end{tabular}

\begin{flushleft}
    \textsc{Note.}\\
     The second column gives measurements properties which are collected from the corresponding references. The third and fourth columns list all the observed properties and fiducial sound horizon scale. The intention of data selection and methods of calculation can be found in Section \ref{sec:BAO}. In the table, $D_V$ is defined in eq.(\ref{equ:DV}), $D_M$ is comoving distance which equals to $D_C$ defined in section \ref{sec:Model}, $c$ is the speed of light, and $H(z)$ is the Hubble parameter.\\
    \vspace{1em}
\end{flushleft}
\end{table*}

    \begin{table*}
        \renewcommand{\arraystretch}{1.5}
        \addtolength{\tabcolsep}{+3pt}
        \centering
        \caption{Fitting results.}
        \label{tab:parameters}
        \begin{tabular}{clcccccc}
            \hline
            Dataset                        & Model              & $\Omega_{m}$            & $\Omega_{k}$            & $n$                     & $f_{e}$                 & $h_0$                   & $r_{*}$                 \\ \hline
            \multirow{4}{*}{SL}            & Flat-$\Lambda$CDM  & $0.685^{+0.204}_{-0.220}$ & ---                     & ---                     & $1.051^{+0.012}_{-0.014}$ & ---                     & ---                     \\
            & Curve-$\Lambda$CDM & $0.183^{+0.247}_{-0.132}$ & $0.160^{+0.143}_{-0.257}$ & ---                     & $1.027^{+0.008}_{-0.008}$ & ---                     & ---                     \\
            & PL                 & ---                     & ---                     & $0.658^{+0.103}_{-0.075}$ & $1.068^{+0.014}_{-0.014}$ & ---                     & ---                     \\
            & RhCT               & ---                     & ---                     & ---                     & $1.036^{+0.007}_{-0.007}$ & ---                     & ---                     \\ \hline
            \multirow{4}{*}{SL+$H(z)$}     & Flat-$\Lambda$CDM  & $0.359^{+0.067}_{-0.057}$ & ---                     & ---                     & $1.031^{+0.008}_{-0.008}$ & $0.664^{+0.030}_{-0.030}$ & ---                     \\
            & Curve-$\Lambda$CDM & $0.291^{+0.089}_{-0.079}$ & $0.057^{+0.098}_{-0.115}$ & ---                     & $1.028^{+0.007}_{-0.007}$ & $0.683^{+0.025}_{-0.025}$ & ---                     \\
            & PL                 & ---                     & ---                     & $0.928^{+0.075}_{-0.063}$ & $1.041^{+0.008}_{-0.008}$ & $0.603^{+0.025}_{-0.025}$ & ---                     \\
            & RhCT               & ---                     & ---                     & ---                     & $1.036^{+0.007}_{-0.007}$ & $0.623^{+0.014}_{-0.014}$ & ---                     \\ \hline
            \multirow{4}{*}{\makecell{SL+$H(z)$+BAO \\ case 1}} & Flat-$\Lambda$CDM  & $0.302^{+0.018}_{-0.017}$ & ---                      & ---                     & $1.026^{+0.007}_{-0.007}$ & $0.689^{+0.018}_{-0.018}$ & $1.463^{+0.036}_{-0.034}$ \\
            & Curve-$\Lambda$CDM & $0.295^{+0.024}_{-0.024}$ & $0.034^{+0.091}_{-0.084}$ & ---                     & $1.028^{+0.008}_{-0.008}$ & $0.684^{+0.021}_{-0.021}$ & $1.466^{+0.037}_{-0.035}$ \\
            & PL                 & ---                     & ---                     & $0.931^{+0.018}_{-0.018}$ & $1.041^{+0.007}_{-0.007}$ & $0.604^{+0.015}_{-0.015}$ & $1.496^{+0.036}_{-0.035}$ \\
            & RhCT               & ---                     & ---                     & ---                     & $1.036^{+0.007}_{-0.007}$ & $0.623^{+0.014}_{-0.014}$ & $1.494^{+0.036}_{-0.035}$ \\ \hline
            \multirow{4}{*}{\makecell{SL+$H(z)$+BAO \\ case 2} }& Flat-$\Lambda$CDM  & $0.302^{+0.018}_{-0.017}$ & ---                     & ---                     & $1.026^{+0.007}_{-0.007}$ & $0.684^{+0.009}_{-0.009}$ & $1.474^{+0.007}_{-0.007}$ \\
            & Curve-$\Lambda$CDM & $0.295^{+0.024}_{-0.024}$ & $0.040^{+0.088}_{-0.083}$ & ---                     & $1.028^{+0.008}_{-0.008}$ & $0.681^{+0.011}_{-0.011}$ & $1.474^{+0.007}_{-0.007}$ \\
            & PL                 & ---                     & ---                     & $0.931^{+0.018}_{-0.018}$ & $1.041^{+0.007}_{-0.007}$ & $0.612^{+0.007}_{-0.007}$ & $1.475^{+0.007}_{-0.007}$ \\
            & RhCT               & ---                     & ---                     & ---                     & $1.036^{+0.007}_{-0.007}$ & $0.631^{+0.005}_{-0.005}$ & $1.475^{+0.007}_{-0.007}$ \\ \hline
                \multirow{4}{*}{\makecell{SL+$H(z)$+BAO \\ case 3} }& Flat-$\Lambda$CDM  & $0.284^{+0.015}_{-0.015}$ & ---                     & ---                     & $1.024^{+0.007}_{-0.007}$ & $0.696^{+0.008}_{-0.008}$ & $1.470^{+0.007}_{-0.007}$ \\
            & Curve-$\Lambda$CDM & $0.292^{+0.022}_{-0.022}$ & $-0.038^{+0.074}_{-0.072}$ & ---                     & $1.022^{+0.008}_{-0.008}$ & $0.699^{+0.009}_{-0.009}$ & $1.470^{+0.007}_{-0.007}$ \\
            & PL                 & ---                     & ---                     & $0.970^{+0.019}_{-0.018}$ & $1.038^{+0.007}_{-0.007}$ & $0.632^{+0.007}_{-0.007}$ & $1.467^{+0.007}_{-0.007}$ \\
            & RhCT               & ---                     & ---                     & ---                     & $1.036^{+0.007}_{-0.007}$ & $0.639^{+0.005}_{-0.005}$ & $1.468^{+0.007}_{-0.007}$ \\ \hline
            \multirow{4}{*}{\makecell{SL+$H(z)$+BAO \\ case 4} }& Flat-$\Lambda$CDM  & $0.307^{+0.015}_{-0.014}$ & ---                     & ---                     & $1.026^{+0.007}_{-0.007}$ & $0.681^{+0.006}_{-0.006}$ & $1.474^{+0.007}_{-0.007}$ \\
            & Curve-$\Lambda$CDM & $0.295^{+0.024}_{-0.024}$ & $0.047^{+0.076}_{-0.074}$ & ---                     & $1.028^{+0.007}_{-0.007}$ & $0.679^{+0.007}_{-0.007}$ & $1.474^{+0.007}_{-0.007}$ \\
            & PL                 & ---                     & ---                     & $0.982^{+0.018}_{-0.017}$ & $1.037^{+0.007}_{-0.007}$ & $0.638^{+0.006}_{-0.006}$ & $1.465^{+0.007}_{-0.007}$ \\
            & RhCT               & ---                     & ---                     & ---                     & $1.036^{+0.007}_{-0.007}$ & $0.641^{+0.004}_{-0.004}$ & $1.466^{+0.007}_{-0.007}$ \\ \hline
        \end{tabular}
        \vspace{1em}
        \begin{flushleft}
            \textsc{Note.}\\ Fitting results from SL+$H(z)$+BAO with flat or Gaussian prior of $r_d$ respectively. In the first column, case 1 represents the flat prior of $r_d$ ($P(r_{*})=U[1,2]$), and case 2 represents the Gaussian prior of $r_d$ ($r_{*} = 1.474 \pm 0.007$). Case 3 represents Gaussian priors on $r_d$ and $H_{0} = 73.52 \pm 1.62\, \rm{km\,s^{-1}\,Mpc^{-1}}$, while case 4 represents Gaussian priors on $r_d$ and $H_{0} = 67.80 \pm 0.09\,\rm{km\,s^{-1}\,Mpc^{-1}}$.
            \vspace{1.5em}
        \end{flushleft}
    \end{table*}

    \begin{table*}
        \renewcommand{\arraystretch}{1.5}
        \addtolength{\tabcolsep}{+10pt}
        \begin{center}
            \caption{Model comparison}
            \label{tab:IC}
            \begin{tabular}{clcc}
                \hline
                Dataset                        & Model              & $\ln E$              & $\ln B$            \\ \hline
                \multirow{4}{*}{SL}            & Flat-$\Lambda$CDM  & ${-202.28}\pm{0.01}$ & ---                \\
                & Curve-$\Lambda$CDM & ${-203.28}\pm{0.01}$ & ${1.00}\pm{0.01}$  \\
                & PL                 & ${-204.14}\pm{0.01}$ & ${1.86}\pm{0.01}$  \\
                & RhCT               & ${-205.76}\pm{0.01}$ & ${3.48}\pm{0.01}$  \\ \hline
                \multirow{4}{*}{SL+$H(z)$}     & Flat-$\Lambda$CDM  & ${-214.53}\pm{0.01}$ & ---                \\
                & Curve-$\Lambda$CDM & ${-214.33}\pm{0.01}$ & ${-0.20}\pm{0.02}$ \\
                & PL                 & ${-218.50}\pm{0.01}$ & ${3.97}\pm{0.02}$  \\
                & RhCT               & ${-216.70}\pm{0.01}$ & ${2.17}\pm{0.02}$  \\ \hline
                \multirow{4}{*}{\makecell{SL+$H(z)$+BAO \\ case 1}} & Flat-$\Lambda$CDM  & ${-226.78}\pm{0.02}$ & ---                \\
                & Curve-$\Lambda$CDM & ${-228.23}\pm{0.02}$ & ${1.45}\pm{0.03}$  \\
                & PL                 & ${-234.11}\pm{0.02}$ & ${7.33}\pm{0.03}$  \\
                & RhCT               & ${-236.97}\pm{0.02}$ & ${10.18}\pm{0.03}$ \\ \hline
                \multirow{4}{*}{\makecell{SL+$H(z)$+BAO \\ case 2}} & Flat-$\Lambda$CDM  & ${-224.43}\pm{0.02}$ & ---                \\
                & Curve-$\Lambda$CDM & ${-225.90}\pm{0.02}$ & ${1.46}\pm{0.03}$  \\
                & PL                 & ${-231.89}\pm{0.02}$ & ${7.46}\pm{0.03}$  \\
                & RhCT               & ${-234.70}\pm{0.02}$ & ${10.27}\pm{0.02}$ \\ \hline
                    \multirow{4}{*}{\makecell{SL+$H(z)$+BAO \\ case 3}} & Flat-$\Lambda$CDM  & ${-225.83}\pm{0.02}$ & ---                \\
                & Curve-$\Lambda$CDM & ${-227.30}\pm{0.02}$ & ${1.47}\pm{0.03}$  \\
                & PL                 & ${-245.24}\pm{0.02}$ & ${19.41}\pm{0.03}$  \\
                & RhCT               & ${-245.31}\pm{0.02}$ & ${19.48}\pm{0.03}$ \\ \hline
                    \multirow{4}{*}{\makecell{SL+$H(z)$+BAO \\ case 4}} & Flat-$\Lambda$CDM  & ${-221.82}\pm{0.02}$ & ---                \\
                & Curve-$\Lambda$CDM & ${-223.28}\pm{0.02}$ & ${1.47}\pm{0.02}$  \\
                & PL                 & ${-245.69}\pm{0.02}$ & ${23.87}\pm{0.03}$  \\
                & RhCT               & ${-242.37}\pm{0.02}$ & ${20.56}\pm{0.03}$ \\ \hline
            \end{tabular}
        \end{center}
    \begin{flushleft}
        \textsc{Note.}\\
Columns $\ln E$ and $\ln B$
        stand for Beyesian evidence and Bayes factor in natural logarithm, respectively. We fix flat-$\Lambda$CDM as a criterion model.
        Others should be compared with flat-$\Lambda$CDM to obtain
        Bayes factor in natural logarithm $\ln B$. Different cases are the same as Table \ref{tab:parameters}. 
        \\
        \vspace{1.5em}
        \end{flushleft}
    \end{table*}

\newpage

\end{document}